\documentclass[12pt,preprint]{aastex}
\newcommand{\be}{\begin{equation}}
\newcommand{\ee}{\end{equation}}

\newcommand{\msun}{{\rm M}_{\sun}}

\begin{document}

\title{On the origin of X-ray emission in some FR Is: ADAF or jet? }
\author{Qingwen Wu\altaffilmark{1,2,3}, Feng Yuan\altaffilmark{1,2} and Xinwu Cao\altaffilmark{1,2}}

\altaffiltext{1}{Shanghai Astronomical Observatory, Chinese Academy of
Sciences, Shanghai, 200030, China; qwwu@shao.ac.cn, fyuan@shao.ac.cn,
cxw@shao.ac.cn}

\altaffiltext{2}{Joint Institute for Galaxy and Cosmology (JOINGC) of SHAO
and USMC, 80 Randan Road, Shanghai 200030, China}

\altaffiltext{3}{Graduate School of Chinese Academy of Sciences, Beijing,
100039, China}

\begin{abstract}
We investigate the X-ray origin in FR Is using the radio,
submillimetre, optical, and {\em Chandra} X-ray data of a small
sample consisting of eight FR I sources. These sources are very dim,
with X-ray luminosities $L_{\rm X}/L_{\rm Edd} \sim 10^{-4}-10^{-8}$
($L_{\rm X}$ is the X-ray luminosity between 2-10 keV). We try to
fit the multiwaveband spectrum using a coupled accretion-jet model.
In this model, the accretion flow is described by an
advection-dominated accretion flow (ADAF) while in the innermost
region of ADAF a fraction of accretion flow is transferred into the
vertical direction and forms a jet. We find that X-ray emission in
the source with the highest $L_{\rm X}$ ($\sim 1.8 \times
10^{-4}L_{\rm Edd}$) is from the ADAF. The results for the four
sources with moderate $L_{\rm X}$ ($\sim$ several $\times
10^{-6}L_{\rm Edd}$) are complicated. Two are mainly from the ADAFs,
one from the jet, and the other from the sum of the jet and ADAF.
The X-ray emission in the three least luminous
sources ($L_{\rm X} \lesssim 1.0\times 10^{-6}L_{\rm Edd}$) is dominated by
the jet although for one source it can also be interpreted by the ADAF
since the quality of X-ray data is low. We conclude that
these results roughly
support the predictions of Yuan \& Cui (2005) where they predict that
when the X-ray luminosity of the system is below a critical value,
the X-radiation will not be dominated by the emission from
the ADAF any longer, but by the jet. We also investigate the fuel
supply in these sources. We find that the accretion rates in four
sources among the five in which we can have good constraints to their
accretion rates must be higher than the Bondi rates. This
implies that other fuel supply, such as the gas released by the
stellar population inside the Bondi radius, should be important.

\end{abstract}


\section{Introduction}

FR I radio galaxies (defined by edge-darkened radio structure) have
lower radio power than FR II galaxies (defined by edge-brightened
radio structure due to compact jet terminating hot spots)
\citep{fr74}. What causes the morphological difference between FR I
and FR II radio galaxies is still unclear. Different explanations of
division of FR I and FR II radio galaxies invoke either the
interaction of the jet with the ambient medium or the intrinsic
nuclei properties of accretion and jet formation processes
\citep*[e.g.,][]{bi95,re96a,gk00}. Accretion mode in low power FR Is
may be different from that in powerful FR IIs. There is growing
evidence \citep*[][]{re96b,do04,gl03,mhd03} to suggest that most FR
I type radio galaxy nuclei, except for a few 3C FR Is with obscured
bright nuclei \citep{cr04}, possess advection-dominated accretion
flow (ADAF; or ``radiative inefficient accretion flows''; Narayan \&
Yi 1994; 1995; see Narayan, Mahadevan \& Quataert 1998 and Kato,
Fukue \& Mineshige 1998 for reviews). In fact, we now have strong
observational evidence that ADAFs may be powering various types of
low-luminosity AGNs (LLAGNs; e.g., Fabian \& Rees 1995; Reynolds et
al. 1996; Quataert et al. 1999; Yuan et al. 2002; Ho, Terashima \&
Ulvestad 2003; Ptak et al. 2004; Yuan \& Narayan 2004; Nemmen et al.
2006; see Narayan 2005, Ho 2005, and Yuan 2007 for reviews) and our
Galactic center Sgr A* \citep*[][]{yq03,ys06}, not only FR Is.

Even so, there are still many details which are unclear and require
detailed investigation. One example is the respective contribution
of ADAFs and jets at various wavebands in LLAGNs. For FR Is and more
general LLAGNs, the least controversial nuclear emission is the
radio emission, which is believed to be dominated by the jet
\citep*[e.g.,][]{wu05}. In the optical waveband, {\em Hubble Space
Telescope} (HST) observations show that the optical luminosities of
FR Is correlate linearly with their radio nuclear luminosities very
well, with little scatter. This, together with the high linear
polarization from the nuclear optical emission argues for a
synchrotron (jet) origin for the nuclear optical emission
\citep*[e.g.,][]{ch99,ha99}. The correlation between the optical
nuclear luminosity and radio nuclear luminosity is dual population
for FR Is and LINERs (as well as Seyfert). \citet{ch05,ch06} suggest
that the optical emission of FR Is comes from the jet, while the
optical emission of the LINERs and Seyfert is dominated by the
accretion flows.

Regarding the X-ray emission, while we usually think that the X-ray
emission of LLAGNs comes from the ADAF
\citep*[e.g.,][]{re96b,qe99a}, recently it has been proposed that in
some individual sources the emission from a jet may be responsible
for the observed X-ray emission (e.g., Yuan et al. 2002 for
NGC~4258; Fabbiano et al. 2003 for IC~1459; Pellegrini et al. 2007
for NGC 821; Garcia et al. 2005 for M~31; and references in Yuan \&
Cui 2005). Then an important question is {\em systematically} in
what kind of condition the radiation from the jet will be important
in X-ray band.

There have been several papers toward answering this question.
Almost all are based on the correlation between the radio and X-ray
luminosities of black hole sources: $L_{\rm R}\propto L^{0.6}_{\rm
X}$, where $L_{\rm R}$ is the radio luminosity at 8.6 GHz and
$L_{\rm X}$ is the 2-11 keV X-ray luminosity. Such a correlation was
originally found in the context of the hard state of black hole
X-ray binaries (Corbel et al. 2003; Gallo et al. 2003; but see Xue
\& Cui 2007 and discussion below) and subsequently
extended to including LLAGNs as well (Merloni et al. 2003;
Falcke et al. 2004; Wang et al. 2006; $\rm K\ddot{o}rding$
et al. 2006; Merloni et al. 2006). Such a correlation has sometimes
been used to argue in favor of a common origin from the jet (e.g.,
Markoff et al. 2003; but see Heinz 2004).
However, the correlation does not necessarily imply
a common jet origin, because it is naturally
expected if the mass accretion rate in the ADAF is positively
correlated to the mass loss rate in the jet, as is very likely the
case. On the other hand, the quantitative result of the correlation
does provide us with some important information on the physics of
accretion flow and jet and their relation. For example,
\citet{mhd03} argue that the X-ray emission from low-luminosity
black hole sources as used to establish the correlation is most
likely dominated by the ADAF rather than a standard thin disk or a
jet.

Most sources in \citet{mhd03} sample have $L_{\rm X}/L_{\rm Edd}
\sim 10^{-7}-10^{-1}$. Based on \citet{mhd03} correlation result,
\citet{yc05} investigate how the correlation will change at lower
luminosities. For this purpose, they first use a coupled ADAF-jet
model to explain the observed correlation (within the range of
$L_{\rm X}/L_{\rm Edd} \sim 10^{-7}-10^{-1}$). The X-ray emission is
modeled by thermal Comptonization emission in the ADAF, while the
radio emission is modeled as due to the synchrotron emission in the
jet. To quantitatively explain the correlation, they find that the
ratio of the mass loss rate in the jet to the mass accretion rate in
the ADAF, $\dot{M}_{\rm jet}/\dot{M}$, is not, but not far away
from, a constant with the changing $\dot{M}$. Extrapolating this
ratio to lower $\dot{M}$ or $L_{\rm X}$ ($<10^{-7}L_{\rm Edd}$) and
assuming that the physics of jet remain unchanged at the same time,
they find that the X-ray emission of the system should be dominated
by the jet when $L_{\rm X}$ is lower than a critical value $L_{\rm
X,crit}$ determined by \be {\rm log}\left(\frac{L_{\rm X,
crit}}{L_{\rm Edd}}\right) =-5.356-0.17{\rm log}\left(\frac
{M}{\msun}\right) \ee The physical reason is that with the decrease
of the accretion rate, both the X-ray radiation from the ADAF and
the jet will decrease. The former decreases faster since it is
roughly proportional to $\dot{M}^2$ while the latter to $\dot{M}$.
Below a certain $\dot{M}$ which corresponds to $L_{\rm X,crit}$,
the X-ray emission from the jet will dominate over the ADAF.
 In this low $\dot{M}$ regime, both
the radio and X-ray emissions are from the jet, thus the radio-X-ray
correlation will change, with the correlation index changing from
$\sim 0.6$ to $\sim 1.2$ (see also Heinz 2004). We want to emphasize
that the normalization of the correlation adopted in Yuan \& Cui
(2005) is based on the data of XTE J1118+480. Because of the large
scatter of the correlation (ref. Merloni et al. 2003), and because
of the statistical feature of the correlation, for individual
sources, the exact value of $L_{\rm X,crit}$ could be significantly
different from that estimated in eq. (1). The prediction thus only
has statistical meaning.

Similar to Yuan \& Cui (2005), Fender et al. (2003) also
pointed out the important role of jets when the X-ray luminosity
of the system, $L_{\rm X}$, is very low. They compared the power of
the jets, $P_{\rm jet}$, and $L_{\rm X}$. They found that
when $L_{\rm X}$ is lower than a critical value, $P_{\rm jet}$ is
larger than $L_{\rm X}$. The difference between this work and
Yuan \& Cui (2005) is that the former compared $L_{\rm X}$ with
the {\em total power of the jet} $P_{\rm jet}$ rather than the
{\em emitted X-ray luminosity} from the jet.

The main aim of the present paper is to check the prediction of Yuan
\& Cui (2005). For this purpose we select a small sample of Donato
et al. (2004) since $L_{\rm X}/L_{\rm Edd}$ ranges from $\sim
10^{-4}$ to $\sim 10^{-8}$. We want to check whether the dominance
of X-ray emission will change from an ADAF to a jet when the
luminosity decreases.

Another reason we choose this sample is that we want to investigate
the question of fuel supply.
Observationally we have good estimation to the Bondi accretion rate in all
the sources in this sample. On the other hand, when the X-ray spectrum comes
from the ADAF, we can obtain the required value of mass accretion rate based
on the accretion disk model. We thus can investigate how good the Bondi
accretion rate is as an indicator of the accretion rate.

In \S2 and \S3, we introduce the sample and the coupled ADAF-jet
model respectively. The modeling results are presented in \S4, and
the discussion and summary are presented in \S5 \& 6. Throughout
this paper, we adopt $H_{0}=70\ \rm km\ s^{-1}\ Mpc^{-1}$,
$\Omega_{M}=0.3$, and   $\Omega_{\Lambda}=0.7$.

\section{Sample}
The FR I sample used for the present investigation is selected
from \citet{do04}. The sources in this sample have estimated black hole mass,
Bondi accretion rate, optical, radio, and X-ray nuclear emission.
    There are 9 FR Is in their sample which have compact core X-ray emission
    and have been observed by $Chandra$. We excluded 3C 270 since
    the optical emission may be obscured by its large intrinsic
    column density ($N_{H}\sim10^{22}\rm cm^{-2}$). Therefore, our final
    sample include 8 FR Is\citep[see][for more details]{do04}. We have
    compiled in Table 1 the nuclear luminosity  obtained from literature in
    the radio, submillimetre, infrared, optical, ultraviolet(UV) and X-ray bands.
    This sample is ideal for us to check the prediction of Yuan \& Cui (2005),
    since the range of X-ray luminosities is $\sim
(10^{-4}-10^{-8}) L_{\rm Edd}$

\section{Coupled accretion-jet model}

We briefly describe the ADAF-jet model here. The readers can refer
to Yuan, Cui \& Narayan (2005) for the details. The accretion flow
is described by an ADAF. In past few years, both numerical
simulations (Stone, Pringle, \& Begelman 1999; Hawley \& Balbus
2002; Igumenshchev et al. 2003) and analytical work (Narayan \& Yi
1994, 1995; Blandford \& Begelman 1999; Narayan et al. 2000;
Quataert \& Gruzinov 2000) indicate that probably only a fraction of
the gas that is available at large radius actually accretes onto the
black hole. The rest of the gas is either ejected from the flow or
is prevented from being accreted by convective motions. Following
the proposal due to \citet{bb99}, we can parameterize the radial
variation of the accretion rate with the parameter $p_{\rm w}$,
$\dot{M}=\dot{M}_{\rm out}(R/R_{\rm out})^{p_{\rm w}}$, where
$\dot{M}_{\rm out}$ is the accretion rate at the outer boundary of
the ADAF $R_{\rm out}$. We calculate the global solution of the
ADAF. The viscosity parameter $\alpha$ and magnetic parameter
$\beta$ (defined as ratio of gas to total pressure in the accretion
flow, $\beta=P_{\rm g}/P_{\rm tot}$) are fixed to be $\alpha=0.3$
and $\beta=0.9$. Another parameter is $\delta$, describing the
fraction of the turbulent dissipation which directly heats
electrons. Following Yuan et al. (2006), we use $\delta=0.3$ and
$p_{\rm w}=0.25$ in all our calculations.

The radiative processes we consider include synchrotron,
bremsstrahlung and their Comptonization. We set the outer boundary
of the ADAF at the Bondi radius $R_{B}=2GM_{\rm BH}/c_{s}^{2}$,
$c_{s}=\sqrt{\gamma kT/\mu m_{p}}$ is the adiabatic sound speed of
the gas at the Bondi accretion radius, $T$ is the gas temperature at
that radius, $\mu=0.62$ is the
    mean atomic weight, $m_{p}$ is the proton mass and $\gamma=4/3$
is adiabatic index of the X-ray emitting gas. After the ADAF
structure is obtained, the spectrum of the flow can be
calculated\citep*[e.g.,][]{yq03}.

The jet model adopted in the present paper is based on the internal
shock scenario, widely used in interpreting gamma-ray
burst (GRB) afterglows \citep*[see,][]{sp01,pi99}. A fraction of the
material in the accretion flow is assumed to be transferred into the
vertical direction to form a jet. The jet is assumed to
include equal numbers of protons and electrons. Since the velocity of the
accretion flow is supersonic near the black hole(BH), a standing
shock should occur at the bottom of the jet because of bending. From
the shock jump conditions, we calculate the properties of the
postshock flow, such as electron temperature $T_{\rm e}$.
 The jet is assumed to have
a conical geometry with half-open angle $\phi$ and bulk Lorentz
factor $\Gamma_{j}$ which are independent of the distance from the
BH. The internal shock in the jet should occur as a result of the
collision of shells with different $\Gamma_{j}$, and these shocks
accelerate a small fraction  of the electrons into power-law energy
distribution with index $p$.  We assume that the fraction of
accelerated electrons in the shock is $\xi_{\rm e}$ and fix
$\xi_{\rm e}=10\%$ in our calculations. Following the widely adopted
approach in the study of GRBs, the energy density of accelerated
electrons and amplified magnetic field are described by two free
parameters $\epsilon_{\rm e}$ and $\epsilon_{\rm B}$. Obviously,
$\xi_{\rm e}$ and $\epsilon_{\rm e}$ are not independent. The
half-open angle of the conical jet is assumed to be $\phi=0.1$,
which is the typical value for the inner jet in FR Is and does not
affect our results \citep*[e.g.,][]{la02}. We find that the Lorentz
factor has only a modest effect on the best fit of the jet spectrum
if it is in the typical range $\Gamma_{j}\sim2-5$ (or $v/c\sim0.9$)
of FR Is \citep*[e.g.,][]{vk02,la02,bo00}. \citet{la99} also
concluded that the typical on-axis velocity of the inner jet was
$v/c\sim0.9$ for FR Is. For simplicity, we set the $\Gamma_{j}=2.3$
(corresponding to $v/c=0.9$) in jet spectra calculations if there is
no estimation on the Lorentz factor.

We consider only synchrotron emission in jet spectrum calculation,
since Compton scattering is probably not important in these FR Is
for several reasons. Firstly, our calculation show that the
synchrotron self-Compton (SSC) in the jet is several magnitude less
than the synchrotron emission in these FR Is since the ratio of the
photon energy density to the magnetic field energy density is very
low(thin long-dashed lines in Figures in this paper). Secondly,
Compton scattering of the external photons from the disk and
emission lines should also not be important because the disk
emission is rather low, and there may be a lack of broad emission
lines in FR Is \citep*[e.g.,][]{ch99}. Thirdly, the inverse Compton
scattering of cosmic microwave radiation should be unimportant for
these FR Is since this mechanism requires high-velocity bulk motion
of the jet, which may be present only in powerful FR II radio
galaxies \citep*[e.g.,][]{ce01}.

We treat the mass loss rate into the jet, $\dot{M}_{\rm
jet}$, and accretion rate in the accretion flow, $\dot{M}_{\rm out}$,
as free parameters when fitting the spectrum, since the jet
formation mechanism is still unclear.
The other free parameters in
spectral fitting  are the electron energy spectral index, $p$, the
electron/magnetic energy parameter, $\epsilon_{\rm e}$,
$\epsilon_{\rm B}$. We use dimensionless accretion rate
$\dot{m}=\dot{M}/\dot{M}_{\rm Edd}$ throughout the paper.
The consistency between the
ADAF and jet models will be ensured by checking whether the value of
$\dot{M}_{\rm jet}/\dot{M}_{\rm out}$, or more precisely
$\dot{M}_{\rm jet}/\dot{M}(10R_s)$ (see next section), is reasonable.

\section{Spectral fitting results}

We use the ADAF-jet model to fit the spectrum of the sources in our
sample. As we state in \S1, the radio emission, and perhaps optical
as well, is from the jet. Although the jet model is more uncertain
than the ADAF, based on the assumption to the jet model described in
\S3, the contribution of the jet to the X-ray band is well
constrained once we require the jet to explain the radio and optical
spectrum. We then adjust the parameter of the ADAF and combine it
with the jet contribution to fit the X-ray spectrum.

\subsection{3C 346}

The radio morphology and power of 3C 346 would rank it as either a
low-power FR II source or a high-power FR I \citep*[e.g.,][]{sp91}.
\citet{co95} used the Very Long Baseline Interferometry (VLBI) radio
core dominance and jet to counter jet ratio to argue for a viewing
angle of the jet is about $<32^{\rm o}$ and a speed of $>0.8c$.
$Chandra$ observation have detected an unresolved core with 2-10 keV
luminosity of $1.9\times10^{43}\ \rm erg\ s^{-1}$ and photon index
of $\Gamma=1.69^{+0.09}_{-0.09}$\citep{do04}. An X-ray knot is also
detected by $Chandra$ with a photon index $\Gamma=2.0\pm0.3$ and no
intrinsic absorption, which roughly corresponding to the brightest
radio and optical knot \citep{wb05}.

Figure 1 shows the fitting result. The dashed, dot-dashed, and the
solid lines show the emission of the jet, ADAF, and their sum
respectively. The parameters of the jet are $\dot{m}_{\rm
jet}=3.5\times10^{-5}$, $\epsilon_{\rm e}=0.14$, $\epsilon_{\rm
B}=0.02$, and $p=2.4$. We find that the jet model can describe well
the nuclear radio and optical emission. But the X-ray emission of
the jet model is several times lower than the $Chandra$
observations. The X-ray emission can be well fitted by the
underlying ADAF. The required accretion rate is $\dot{m}_{\rm
out}=2.8\times10^{-2}$. The ratio of mass loss rate in the jet to
accretion rate of ADAF at 10 $R_{\rm S}$ is about $\dot{m}_{\rm
jet}/\dot{m}(10 R_{\rm S})=$0.9\%, where $R_{\rm S}$ is
Schwarzschild radius.

This source is relatively luminous, with $L_{\rm X}/L_{\rm
Edd}=1.8\times10^{-4}$. The X-ray luminosity is well above the
critical luminosity defined in eq. (1). Thus according to Yuan \&
Cui (2005), the X-ray emission should be dominated by the ADAF
rather than the jet. Our modeling result confirms this prediction.

\subsection{B2 0755+37}

  B2 0755+37 is a well studied nearby FR I ($z=0.0428$) with two
  large symmetrical lobes and very asymmetric jets. The inferred
  velocity of the jet is $\sim 0.9c$ (1 arcsec from the core, $\sim0.5$ kpc),
   and the viewing angle is about $30^{\rm o}$ \citep*[][]{bo00}. In the optical band,
  the galaxy has a rather smooth appearance
  without any sign of dust obscuration, and a  bright optical nucleus is seen
  at its center \citep{ca02}. The optical jet is also detected, and
  the optical brightness profile is similar to that of radio
\citep{pa03}. A power-law fit to the nuclear emission observed by
$Chandra$ results in a power-law index $\Gamma=2.18^{+0.28}_{-0.19}$ and
2-10 keV luminosity of $6.1\times10^{41}\ \rm erg\ s^{-1}$
($\sim5.2\times10^{-6}L_{\rm Edd}$) \citep{do04}.

Figure 2 shows the fitting result of B2 0755+37. The dashed-line
shows the jet emission. The parameters are $\dot{m}_{\rm
jet}=1.75\times10^{-5}$, $\epsilon_{\rm e}=0.12$, $\epsilon_{\rm
B}=0.01$, and $p=2.23$. We find that the jet model can not only well
describe the radio and optical spectra, but also X-ray spectrum.

The dot-dashed line shows the emission from an ADAF model, with
$\dot{m}_{\rm out}=8.9\times10^{-3}$. This is of course not a `fit'.
We show this result to illustrate that an ADAF would predict a much
harder X-ray spectrum than observed. This is another evidence for
the dominance of the X-ray emission by the jet. Obviously
$\dot{m}_{\rm out}$ should be the upper limit of accretion rate.

\subsection{3C 31}

    3C 31 is also a twin-jet FR I radio galaxy (redshift
    $z$=0.0169), hosted by the D galaxy NGC 383.  3C 31 has been
    widely studied in the radio, optical and X-ray bands
    \citep*[e.g.,][]{ma00,la02,ha02}. High quality radio imaging
    with Very Large Array (VLA) allowed \citet{la02} to make
    detailed models of velocity field in the jets within 30 arcsec
    of the nucleus, on the assumption that the jets are
    intrinsically symmetrical and anti-parallel. They inferred the
    angle to the line of sight to be $52^{\rm o}$ with an uncertainty of
    a few degrees, and found that central velocity is $\sim$0.87$c$ for the inner region (0 to 1.1
    $kpc$). The $HST$ image reveals a nearly
    face-on dust disk surrounding the unresolved galaxy's nucleus \citep*[e.g.,][]{ma99}.
   From the $Chandra$ observation of November 2000,  the
   inner region has been resolved in the point-like X-ray core and an
   extended X-ray jet \citep{ha02}. The spectrum of the core is quite flat, with
   photon index $\Gamma=1.48^{+0.28}_{-0.32}$ and the 2-10 keV nuclear
   X-ray luminosity is $4.7\times10^{40}\rm erg\ s^{-1}$ ($\sim4.4\times10^{-6}L_{\rm Edd}$)\citep{ev06}.
   The X-ray jet can be fitted by a power-law model with a photon index
   $\Gamma =2.09\pm0.16$ \citep*[][]{ha02}.

Figure 3 shows the fitting result of 3C 31. The dashed and
dot-dashed lines are for the emissions from the jet and the ADAF,
respectively, and the solid line shows their sum. The parameters of the jet
are $\dot{m}_{\rm jet}=2.7\times10^{-5}$, $\epsilon_{\rm e}=0.2$,
$\epsilon_{\rm B}=0.02$, and $p=2.5$. We find that the radio,
optical and even the soft X-ray nuclear emission (e.g., 1 keV) can
be well fitted by a pure jet model. However, the hard X-ray cannot
be fitted by a jet, but can be well fitted by the underlying ADAF
with accretion rate $\dot{m}_{\rm out}=3.7\times10^{-3}$. The ratio,
$\dot{m}_{\rm jet}/\dot{m}(10\ R_{\rm S})$, is about 9\%.

\subsection{3C 317}

The radio galaxy 3C 317 is associated with CD galaxy UGC 9799 ($z=0.0345$),
located at the centre of the cooling flow cluster A 2052. The
radio observation shows that it may be a very young radio
source \citep{ve04}.  Unresolved compact core is observed by
$HST$ in both optical and UV band. However, the core emission
show strong variability in the UV band between 1994 and 1999 \citep*[by a factor of
10,][]{ch02}. The spectral index of optical-UV is very large,
$\alpha_{\rm opt-UV}=3.3$ \citep*[$F_{\nu}\propto\nu^{-\alpha}$,][]{ch02}.
The central compact core is also detected by $Chandra$ with
2-10 keV power-law luminosity $2.97\times10^{41}\rm erg\
s^{-1}$ ($\sim3.4\times10^{-6}L_{\rm Edd}$) and a photon index
$\Gamma=1.81^{+0.13}_{-0.1}$ \citep{do04}.

Figure 4 shows the fitting result of 3C 317. The dashed, dot-dashed
lines show the emissions from the jet and ADAF respectively, while
the solid line shows their sum. The parameters of the jet are
$\dot{m}_{\rm jet}=1.7\times10^{-5}$, $\epsilon_{\rm e}=0.2$,
$\epsilon_{\rm B}=0.15$, and $p=2.25$. It is difficult to fit the
radio and optical emission simultaneously due to the steep
optical-UV spectrum. Since the UV flux is very sensitive to the
extinction, such a steep spectrum may be not intrinsic.  We
therefore use $R$ band (F702W) in our fits, which is less
susceptible to extinction than the UV band and less contaminated by
the possible dust emission than the near-infrared band($H$ band,
F160W). The radio, optical and even soft X-ray band (e.g., 1 keV)
emission can be fitted by a jet model. The hard X-ray spectrum can
not be well fitted by the pure jet or underlying ADAF alone. Rather,
it can be well fitted by their sum. The required accretion rate of
the ADAF is $\dot{m}_{\rm out}=4.7\times10^{-3}$.

\subsection{B2 0055+30}

The source B2 0055+30 (NGC 315) is associated with a giant
elliptical galaxy at a redshift of 0.0168, which has a two-side
structure extending to roughly one degree on the sky
\citep{br79}. \citet{ca05} applied symmetrical, deceleration,
relativistic jet model to fit the radio jet, and derived the
inclination to the line of sight of $38^{\rm o}\pm2^{\rm o}$ and
their on-axis velocity is $\beta=v/c\sim0.9$. The jet is also
detected by $Chandra$ with a power law index $\Gamma=1.5\pm0.7$
\citep{wb03}.  The 2-10 keV nuclear luminosity detected by $Chandra$
is $5.1\times10^{41}\rm erg\ s^{-1}$ ($\sim2.4\times10^{-6}L_{\rm Edd}$)
with a photon index $\Gamma=1.56^{+0.17}_{-0.09}$ \citep{do04}.

Figure 5 shows the fitting result of B2 0055+30.  The dashed and
dot-dashed lines show the emissions from the jet and ADAF
respectively, while the solid line shows their sum. The parameters
of the jet are $\dot{m}_{\rm jet}=7.0\times10^{-6}$, $\epsilon_{\rm
e}=0.05$, $\epsilon_{\rm B}=0.02$, and $p=2.2$. We can see from the
figure that the radio and optical emission can be well fitted by the jet
model. However, the X-ray spectrum is too hard to  be fitted by the
jet, but can be well fitted by the ADAF with the accretion rate
$\dot{m}_{\rm out}=2.7\times10^{-3}$. The ratio $\dot{m}_{\rm
jet}/\dot{m}(10\ R_{\rm S})$ is about 3.5\%.

\subsection{3C 66B}

3C 66B has a redshift of 0.0212 and is associated with a
thirteen-magnitude elliptical galaxy in a small group in the
vicinity of the cluster Abell 347. The two-sided inner radio
jets is seen, with both jets curving toward the east at distance
$>20^{''}-30^{''}$ from the core \citep{le86}. $HST$ observed
the optical jet on scales of $\sim0.1$ arcsec \citep{ma91}. The
jet also been imaged with $Infrared\ Space\ Observatory\ (ISO)$
and $Chandra$ \citep{ta00,ha01}. The multi-wavelength extended
jet emission can be well fitted with the synchrotron emission
\citep*[e.g.,][]{ta00}. The point-like nucleus is detected by $Chandra$
with 2-10 keV luminosity $1.1\times10^{41}\rm ergs^{-1}$
($\sim1\times10^{-6}L_{\rm Edd}$) and a photon index
$\Gamma=2.17^{+0.14}_{-0.15}$ \citep*[e.g.,][]{do04}.

Figure 6 shows the fitting result of nucleus of 3C 66B. We find that
both the radio, submillimetre, optical and X-ray can be fitted by
pure jet model very well (dashed line). The parameters of the jet
are $\dot{m}_{\rm jet}=1\times10^{-5}$, $\epsilon_{\rm e}=0.18$,
$\epsilon_{\rm B}=0.02$, and $p=2.35$. For illustration purpose, we
also show by the dot-dashed line the X-ray emission using an ADAF
model with $\dot{m}_{\rm out}=2.6\times10^{-3}$. We can see that the
predicted spectrum by an ADAF is too hard to be consistent with the
observation (so the accretion rate in the ADAF should be smaller
than $2.6\times10^{-3}$). The X-ray emission in this source should
be from the jet.

\subsection{3C 449}

3C 449 is a low-redshift ($z$=0.0171) twin-jet FR I type radio
galaxy which hosted by the elliptical galaxy UGC 12064. Its symmetrical inner jets have
been well studied in the radio observation \citep*[e.g.,][]{fe99}. On
large scales, the southern jet flares into a lobe, while the
northern jet continue to be well collimated until it fades into the noise
\citep*[e.g.,][]{an92,fe99}. From the application of the
adiabatic model to the jet, evidence of a strong jet
deceleration within $10^{''}$ (5 kpc) from the nucleus is found.
A satisfactory fit to the data is found assuming an initial jet
velocity 0.9$c$, and a jet inclination to the line of sight of
$82.5^{\rm o}$ \citep*[e.g.,][]{fe99}. The nucleus is an unresolved point source
from $HST$ observation \citep*[e.g.,][]{ma99}.
$Chandra$ observation exhibit a X-ray central compact core with a
power-law photon index  $\Gamma=1.67^{+0.45}_{-0.49}$ and the
2-10 keV power-law luminosity is less than $2.9\times10^{40}\rm erg\ s^{-1}$
($\sim8\times10^{-7}L_{\rm Edd}$) \citep*[][]{ev06}. \citet{ev06} shown that the X-ray
luminosity observed by Chandra is several times lower than that
measured by $XMM-Newton$ and it's photon index is also flatter than
$\Gamma=2.13^{+0.65}_{-0.55}$ of $XMM-Newton$  \citep*[][]{do04}.
Although the variability is a possible  explanation for the
difference observed by $Chandra$ and $XMM-Newton$, it may be also due to
the different resolution of the telescopes. The spectrum is
extracted in a smaller radius circle (typically 2.5 pixels or $1^{''}.23$ in
the case of $Chandra$) than $35^{''}$ in the case of {\em XMM-Newton} \citep{ev06}.
$Chandra$ observation should reflect the intrinsic nuclear X-ray
emission, while {\em XMM-Newton} observation may be contaminated by the X-ray jet
emission and/or X-ray binaries.

Figure 7 shows the fitting result of 3C 449.  We find that the
radio, optical and X-ray emissions can be roughly fitted by a pure jet
model shown by the thick long-dashed line. The parameters are $\dot{m}_{\rm
jet}=4.0\times10^{-5}$, $\epsilon_{\rm e}=0.45$, $\epsilon_{\rm
B}=0.003$, and $p=2.25$.

Since the observational error in the X-ray photon index is very
large, we also try to fit the X-ray emission with the sum of a jet
and an ADAF. The thin short-dashed and dot-dashed lines show the
emissions from the jet and the ADAF respectively and the solid line
shows their sum. The parameters of the jet are $\dot{m}_{\rm
jet}=2.0\times10^{-5}$, $\epsilon_{\rm e}=0.35$, $\epsilon_{\rm
B}=0.01$, and $p=2.4$. The accretion rate of the ADAF is
$\dot{m}_{\rm out} =1.9\times10^{-3}$. We can see that the X-ray
emission can also be (slightly better because of more free
parameters) fitted by the sum of ADAF and jet. The value of
$\dot{m}_{\rm jet}/\dot{m}(10\ R_{\rm s})$ is about 14.3\%.
Higher quality X-ray data is desired to further constrain
the origin of the X-ray emission in 3C 449.

\subsection{3C 272.1}

3C 272.1 (M 84=NGC 4374) is an E1 galaxy in the core of the Virgo
Cluster ($z$=0.0029). Radio observations at 1.4 and 4.9 GHz show two
lobes and a jet \citep{la87}. In the SCUBA 850-$\mu\rm m$
submillimetre image, the galaxy is found to be a point source
(diameter $<$15 arcsec, 1.5kpc). The submillimetre emission was
suggested to be from the inner jet, or from the emission from the
thermal emission of cold diffuse dust \citep{le00}. {\em HST} show that
the optical-to-UV continuum is very red, similar to the spectral
energy distribution of BL Lac \citep{bg00}. The $Chandra$ image show
that the soft X-ray emission have a very disturbed morphology.
\citet{do04} defined this source as a candidate compact core X-ray
source, since that its radial profile cannot be fitted with a
$\beta$-model. The 2-10 keV luminosity is $2.2\times10^{39}\rm
ergs^{-1}$ ($\sim6.8\times10^{-8}L_{\rm Edd}$) with a photon index
$\Gamma=2.14^{+0.34}_{-0.30}$ \citep{ev06}.

Figure 8 shows the fitting result of nucleus of 3C 272.1. The dashed
and the dot-dashed lines show the emissions from the jet and the
ADAF respectively. The parameters of the jet are
$\dot{m}_{\rm jet}=4.9\times10^{-6}$, $\epsilon_{\rm
e}=0.28$, $\epsilon_{\rm B}=0.005$, and $p=2.5$. We can see
that the radio, submillimetre, optical and especially X-ray emission
can be fitted by the jet model very well. On the other hand, the predicted
spectrum by an ADAF (with $\dot{m}_{\rm out}=1.9\times10^{-3}$)
is too hard to be consistent with observation. So in this source, the X-ray emission
is dominated by a jet.

\section{Discussion}

In this paper we have investigated the origin of the X-ray emission
in a small sample from FR Is in Donato et al. (2004). The accretion
flow in FR Is is generally believed to be described by an ADAF
rather than a standard thin disk since the Bondi accretion rates
(which should be a lower limit, see discussion below) would produce
a luminosity several orders of magnitude higher than that observed
if we assume a standard thin disk with the efficiency of $\sim 0.1$
as in a standard thin disk. Two possibilities exist for the X-ray
origin in FR Is--ADAF or jet. We use a coupled ADAF-jet model to fit
the multiwaveband spectrum to try to investigate this problem. More
specifically, we want to examine the prediction of Yuan \& Cui
(2005) that statistically the X-ray emission in LLAGNs should be
dominated by ADAFs when their X-ray luminosity is higher than $\sim$
a few $\times 10^{-7}L_{\rm Edd}$, but will be dominated by jet when
the luminosity is lower than this value.

We find that the jet can well describe the radio and optical spectra
for all FR Is in our sample except the optical/UV spectrum in
 3C 317 (see our argument on this
point in \S 4.4). The soft X-ray flux at $\sim$ 1 keV of all FR Is
is roughly consistent with the predictions of the jet. This result
indicates why a tight correlation is found among radio, optical and
soft X-ray \citep*[e.g.,][]{ch99,ev06,ba06}.

For the source with the highest luminosity in our sample, 3C 346,
which has $L_{\rm X}=1.8 \times 10^{-4} L_{\rm Edd}$, its X-ray
spectrum is dominated by the ADAF and the jet contribution is
negligible. However, for the four sources with ``intermediate
luminosities'' (B2 0755+37, 3C 31, 3C 317, B2 0055+30; $L_{\rm
X}=(2.4-5.2) \times 10^{-6} L_{\rm Edd}$), their X-ray origin is
complicated. The X-ray spectra of B2 0755+37 is dominated by the jet
while for 3C 31 and B2 0055+30 they are dominated by the ADAF. For
the other one (3C 317), the contributions of the ADAF and jet are
comparable. For the three least luminous sources (3C 66B, 3C 449 and
3C 272.1) which have $L_{\rm X}=(6.8\times 10^{-8}-1\times 10^{-6})
L_{\rm Edd}$, their X-ray spectra are dominated by the jet. The
X-ray emission of 3C 449 is also interpreted by the sum of a jet and
an ADAF, which requires higher quality data to further constrain it.

Our results are roughly consistent with the predictions of
Yuan \& Cui (2005). The ``intermediate luminosity'' here in our
sample corresponds to the critical luminosity in Yuan \& Cui (2005).
However, the former is about 10 times higher than the latter. The
value of the critical luminosity depends on the ratio of the mass loss
rate in the jet to the mass accretion rate in the ADAF. This ratio
is adopted in Yuan \& Cui (2005) from fitting the data of a black hole
X-ray binary---XTE J1118+480. The current result indicates that the ratio
in FR Is is about 10 times higher than in XTE J1118+480.
This seems reasonable given that the jet in FR Is is
systematically more powerful than in normal LLAGNs. One possible reason
could be that systematically the black holes in FR Is are spinning
more rapidly.

The sample we used in the present paper is rather small. Obviously
to systematically study the X-ray origin of FR Is and more generally
LLAGNs a much larger sample is required and this is our future work
(Yuan et al. in preparation). We note that current results from
literature seem to support the prediction of Yuan \& Cui (2005). In
addition to the observational evidences listed in Yuan \& Cui
(2005), the X-ray emission of some LLAGNs claimed to be dominated by
jet always have $L_{\rm X} < L_{\rm X, crit}\sim (10^{-6}-
10^{-7})L_{\rm Edd}$---NGC 821: $L_{X}/L_{\rm Edd}\sim
3.0\times10^{-7}$ (Fabbiano et al. 2004); IC 1459: $L_{X}/L_{\rm
Edd}\sim 2.9\times10^{-7}$ (Fabbiano et al. 2003); NGC 4594:
$L_{X}/L_{\rm Edd}\sim 1.0\times10^{-7}$ (Pellegrini et al. 2003);
M~31: $L_{X}/L_{\rm Edd}\sim 2.2\times10^{-10}$ (Garcia et al.
2005). On the other hand, the ADAF contribution usually dominates
over jet for the sources with $L_{\rm X}>L_{\rm X, crit}$---NGC
4261: $L_{X}/L_{\rm Edd}\sim 1.4\times10^{-6}$ (Gliozzi et al.
2003); NGC 3998: $L_{X}/L_{\rm Edd}\sim2.0\times10^{-6}$ (Ptak et
al. 2004). Another interesting result we would like to mention
is the correlation between $\dot{m}(10R_s)$ and
$\dot{m}_{\rm jet}/\dot{m}(10R_s)$, i.e., columns 7 \& 8 in Table 2.
We can find that the sources with the highest $\dot{m}(10R_s)$ have the smallest
$\dot{m}_{\rm jet}/\dot{m}(10 R_{S})$ (3C 346), and lower $\dot{m}(10R_s)$
sources have larger $\dot{m}_{jet}/\dot{m}(10 R_{S})$. This result is
roughly consistent with the prediction of Yuan \& Cui 2005 (their Fig. 2).
But we want to emphasize again that a larger sample is required to
more seriously check this prediction.

As we emphasize earlier, the radio-X-ray correlation and therefore
the prediction of Yuan \& Cui (2005) only holds in statistical
sense and may not be valid for any individual source. In fact, a most recent critical examination of the radio/X-ray data
in a sample of black hole X-ray binaries has shown that the correlation
is very diverse and some sources don't show such a correlation
(Xue \& Cui 2007; see also Rodriguez et al. 2007).
Recently Gallo et al. (2006) obtain one
radio/X-ray data set in the quiescent state of A0620-00,
with $L_{\rm X}\sim 10^{-8.5}L_{\rm Edd}\ll
L_{\rm X, crit}$. They find that the data lies on the extrapolation
of the radio-X-ray correlation, without change of the correlation index.
Thus A0620-00 may be another exception.
On the other hand, however, we would like to point out a caveat in Gallo et al.
(2006).  Given the very large scatter in the radio-X-ray correlation
(Merloni et al. 2003; Gallo et al. 2003), it is not appropriate to connect
one data point of {\em a source} with the data of {\em other
different} sources because their normalization may be different.
Rather, one should combine the radio and X-ray
data at different luminosites only for A0620-00.
Although we do have X-ray and radio observations
during its 1975 outburst (Kuulkers 1998), we are unfortunately
not able to convert the X-ray ``counts'' to physical flux due
to instrumental reasons.

Typically Bondi accretion rate is a good estimation to the mass
accretion rate. However, \citet{pe05} show that there is no relation
between the nuclear X-ray luminosity and Bondi accretion rate in
LLAGNs, and X-ray emission of some sources is higher than the values
predicted by ADAFs with Bondi accretion rate. In this paper, the
Bondi accretion rate in our sample have been estimated
\citep*[][]{do04}. We find that the accretion rates $\dot{m}_{\rm
out}$ required in our model of four FR Is (3C 346, 3C 31, 3C 449, 3C
317) among the five in which we can have good constraints to their
accretion rates are higher than their Bondi rates by factors of 9,
18, 112\footnote{The accretion rate of 3C 449 is calculated from the
X-ray emission based on the ADAF model, which should be the upper
limit if the X-ray emission is dominated by the jet as discussed in
\S 4.7 and \S 5.}, 1.05, respectively. Given that the radial
velocity of the accretion flow is $\alpha c_s$, a more accurate
estimation of the accretion rate is $\dot{m}_{\rm
out}\sim\alpha\dot{m}_{\rm Bondi}$ where $\alpha$ is the viscous
parameter \citep{na02}. Therefore the Bondi accretion rate is only a
lower limit of the real rate and other fuel supply must be
important, such as the gas released by the stellar population inside
the Bondi radius (Soria et al. 2006; Pellegrini 2007).
In our calculation, given the
theoretical uncertainties about the values of $p_w$ and $\delta$,
we choose the values of these two parameters from the best studied
source, Sgr A* (Yuan et al. 2006; Yuan, Quatatert \& Narayan 2003),
because we think the physics of accretion should be the same, independent
of various sources. A higher $\delta$ and
lower $p_w$ incline to require a smaller accretion rate.
But we find that even though we use $\delta=0.5$,
which mean half of the viscous dissipation directly heats electrons,
and $p_w=0$, which mean no outflow, the required accretion rates for
3C 346, 3C 31, and 3C 449 are still larger than their Bondi accretion rates.
Therefore, other fuel supply must be important in these sources.

The kinetic luminosity, $L_{\rm kin}=\Gamma_{j}(\Gamma_{j}-1)
\dot{M}_{\rm jet}c^{2}$, can be derived from our modeling results.
We use $\eta_{\rm jet}=L_{\rm kin}/\dot{M}(10R_{S})c^{2}$ to
describe the efficiency of the jet power converted from the
accretion power, where $\dot{M}(10R_{S})$ is mass accretion rate at
$10R_{S}$. We find that $\eta_{\rm jet}=0.03-0.44$ for the sources
in this sample (see Table 2), and most of them (six of eight) have
$\eta_{\rm jet}$ significantly higher than 0.057 which is the
largest available accretion energy at the innermost stable circular
orbit (ISCO) for a nonrotating black hole. This either imply that
the black holes in these sources are spinning rapidly (so ISCO is
smaller thus more accretion energy is available, or the jet power is
extracted from the spinning black holes via the BZ process;
Blandford \& Znajek 1977; Reynolds et al. 2006), or the accretion
energy within ISCO can be extracted through magnetic field.
We have assumed that the jet includes equal numbers of
protons and electrons. If the jet is enhanced by pairs,
it could give the same emission but with much less kinetic luminosity
and less BH spins.

\section{Summary}

We have fitted the multiwaveband spectra of 8 FR Is ranging from
radio to X-ray with our coupled ADAF-jet model. We find that the
origin of X-ray emission can be from ADAF, jet, or their sum,
depending on the ratio of $L_{X}/L_{\rm Edd}$, here $L_{\rm X}$ is
the X-ray luminosity. When $L_{\rm X}$ is significantly larger than
a critical value $L_{X, \rm crit}$, the X-ray emission will be
dominated by an ADAF. When $L_{\rm X}$ is significantly smaller than
$L_{X,\rm crit}$, it will be dominated by a jet. The contributions
of the ADAFs and jets are capable when $L_{\rm X} \sim L_{X, \rm
crit}$. These results roughly support the prediction of Yuan \& Cui
(2005), except that the value of $L_{\rm X, \rm crit}$ here, several
times of $10^{-6}L_{\rm Edd}$, is $\sim$ 10 times higher than the
predicted value in Yuan \& Cui (2005). This discrepancy may indicate
that the jet in FR Is are systematically stronger than in general
LLAGNs.

\acknowledgments
We thank Wei Cui, Luis Ho, and Rodrigo Nemmen for their
valuable comments, and D. Donato for providing us with his observational
data. This work is supported by the One-Hundred-Talent Program of
China, the National Science Fund for Distinguished Young Scholars
(grant 10325314), and the NSFC (grants number 10333020, 10543002,
and 10543003).

\clearpage
\begin{deluxetable}{lcclc}
\tabletypesize{\scriptsize}
 \tablecaption{Points used in Spectral Energy Distributions } \tablewidth{0pt} \tablehead{
\colhead{Filter} & \colhead{$\rm log10(\nu)$}& \colhead{$\rm
log10(\nu L_{\nu})$}& \colhead{$\rm Resolution^{a}$} &
\colhead{Ref.} } \startdata
                                   $3\rm C\ 346$        \\
\hline

1.7 GHz   &   9.23    &   41.26   &$\sim 1 \rm mas$     & C95 \\
5   GHz   &   9.70    &   41.90   &  $\sim 1^{"}$          & G88\\
8   GHz   &   9.92    &   42.13   &  $\sim 1\ \rm mas$   & C95 \\
F702W     &   14.63   &   43.14   &  $\sim 0.1^{"}$          & C99 \\
\hline\\

               $\rm B2\ 0755+37$          \\
\hline
1.7 GHz   &   9.23    &   40.03   &  $\sim 1^{"}$    & W01 \\
5   GHz   &   9.70    &   40.72   & $\sim 1^{"}$     & C02 \\
F702W     &   14.63   &   42.04   &   $\sim 0.1^{"}$        & C02 \\
\hline\\

     3$\rm C\ 31$        \\

\hline
1.7 GHz   &   9.23    &   38.75    & $\sim 5\ \rm mas$    & X00 \\
5   GHz   &   9.70    &   39.46    &  $\sim 1\ \rm mas$    & G01\\
8.6 GHz   &   9.93    &   39.70    &   $\sim 1^{"}$      & H02\\
345 GHz   &   11.53   &   41.22    &   $\sim 50^{"}$   & Q03 \\
F555W     &   14.57   &   40.86    &   $\sim 0.1^{"}$         & K02 \\
F814W     &   14.73   &   40.85    &   $\sim 0.1^{"}$       & K02 \\
\hline\\

   $3\rm C\ 317$        \\
\hline
1.7 GHz   &   9.23    &   40.19    &   $\sim 1\ \rm mas$   & V00 \\
5   GHz   &   9.70    &   40.73    &   $\sim 1\ \rm mas$   & V00 \\
F210M     &   15.13   &   40.46    &  $\sim 0.1^{"}$          & C02\\
F702W     &   14.63   &   41.37    &   $\sim 0.1^{"}$         & C02\\
F160W     &   14.27   &   41.79    &   $\sim 0.1^{"}$         & T03\\
\hline\\

   $\rm B2\ 0055+30 $        \\
\hline
5   GHz   &   9.70    &   40.26    &   $\sim 1^{"}$   & G05 \\
F814     &    14.57   &   41.18    &  $\sim 0.1^{"}$       & C02\\
\hline\\

   3C 66B       \\
\hline
1.7 GHz   &   9.23    &   39.43    & $\sim 5\ \rm mas$    & X00 \\
5   GHz   &   9.70    &   39.97    & $\sim 1\ \rm mas$    & G01 \\
345 GHz   &   11.54   &   41.50    & $\sim 50^{"}$       & Q03\\
LW1       &   13.82   &   42.54    &  $\sim 1^{"}$   & Q03 \\
LW2       &   13.32   &   42.33    &  $\sim 1^{"}$   & Q03\\
LW3       &   13.40   &   42.54    &  $\sim 1^{"}$   & Q03\\
F814W     &    14.57   &  41.60    &  $\sim0.1^{"}$       & C02\\
\hline\\

        3$\rm C$\ 449\\
\hline
1.5 GHz   &   9.17    &   38.26   &  $\sim 1^{"}$    & S97 \\
5   GHz   &   9.70    &   39.08   &  $\sim 1^{"}$    & S97\\
8.3 GHz   &   9.92    &   39.37   &  $\sim 1^{"}$    & S97\\
F702W     &   14.63   &   40.82   &   $\sim 0.1^{"}$       & C02 \\
\hline\\

 3$\rm C\ 272.1 $        \\
\hline
1.7 GHz   &   9.23    &   37.72    & $\sim 1\ \rm mas$    & J81 \\
5   GHz   &   9.70    &   38.30    &  $\sim 1^{"}$      & G88 \\
8.1 GHz   &   9.90    &   38.48    &$\sim 1\ \rm mas$     & J81 \\
146 GHz   &   11.17   &   39.64    &  $\sim 10^{"}$   & L00\\
221 GHz   &   11.34   &   39.82    &  $\sim 10^{"}$   & L00\\
345 GHz   &   11.53   &   39.96    & $\sim 10^{"}$    & Q03\\
350 GHz   &   11.54   &   40.10    & $\sim 10^{"}$    & L00\\
677 GHz   &   11.82   &   40.16    &  $\sim 10^{"}$   & L00\\
LW3       &   13.32   &   40.31    & $\sim   1^{"}$      & Q03\\
LW7       &   13.49   &   40.57    &    $\sim 1^{"}$     & Q03\\
LW2       &   13.65   &   40.94    &    $\sim 1^{"}$     & Q03\\
L         &   13.93   &   40.53    &  $\sim 1^{"}$       & Q03\\
F205W     &   14.16   &   40.05    &   $\sim 0.1^{"}$      & B00\\
F160W     &   14.27   &   40.16    &   $\sim 0.1^{"}$      & B00\\
F110W     &   14.43   &   40.11    &    $\sim 0.1^{"}$     & B00\\
F814W     &   14.57   &   40.00    &    $\sim 0.1^{"}$     & C02\\
F547W     &   14.73   &   39.98    &    $\sim 0.1^{"}$     & B00\\

\enddata

\tablecomments{The F110W, F160W, F205W filters refer to those from
NICMOS (on board $HST$) $1-2\mu \rm m$. F547W, F702W, F814W filters
refer to optical WFPC2/$HST$ measurements. The LW1, LW2, LW3, LW7
filters refer to $4-15\mu$m data from ISOCAM images. L band
measurements are based on ground based images obtained at the
Infrared Telescope Facility(IRTF, \citet{qu03}).}
 \tablerefs{X00 VLBA \citep{x00}; G01 VLBI \citep{gi01}; H02
VLA \citep{ha02}; Q03 ISO and IRTF \citep{qu03}; K02 HST
\citep{vk02}; V00 VLBI \citep{v00}; C95 VLBI \citet{co95}; C99 HST
\citep{ch99}; C02 HST \citep{ca02}; G88 VLA \citep{g88}; G05 VLA
\citep{gi05}; S97 VLA \citep{ka97}; C02 HST \citep{ca02}; T03 HST
\citep{tr03} L00 SCUBA \citep{le00}; B00 HST \citep{bg00}; J81 VLBI
\citep{jo81}; W01 VLA \citep{wb01};}

\tablenotetext{a}{The resolution of different telescopes which are
listed in references. The `mas' means milli-arcsecond.}
\end{deluxetable}

\clearpage
\begin{deluxetable}{lcccccccccc}
\rotate \tabletypesize{\scriptsize}
 \tablecaption{Accretion and jet properties } \tablewidth{0pt}
 \tablehead{\colhead{Source}& \colhead{Redshift} & \colhead{$\rm Angle(deg)^{a}$} &\colhead{$\rm
log10\ M_{\rm BH}^{b}$} & \colhead{$L_{\rm X}/L_{\rm Edd}^{c}$} &
 \colhead{$\dot{m}_{\rm jet}$} &
\colhead{$\dot{m}(10R_{\rm S})$} & \colhead{$ratio^{d}$(\%)} &
\colhead{$\dot{m}(R_{\rm B})^{e}$} & \colhead{$\dot{m}_{B}^{f}$}
&\colhead{$L_{\rm Kin}/\dot{M}(10R_{\rm S})c^{2}$} }

\startdata
3C 346       & 0.1620  & 30   & 8.89   &  $1.8\times10^{-4}$    &  $3.5\times10^{-5}$   &   $4.4\times10^{-3}$     & 0.91       & $2.8\times10^{-2}$   & $3.1\times10^{-3}$  &   0.03 \\
B2 0755+37   & 0.0428  & 34   & 8.93   &  $5.2\times10^{-6}$    &  $1.75\times10^{-5}$  &   $<4.5\times10^{-4}$    & $>3.9$     & $<8.9\times10^{-3}$  & $3.2\times10^{-2}$  &   $>0.10$\\
3C 31        & 0.0170  & 52   & 7.89   &  $4.4\times10^{-6}$    &  $2.7\times10^{-5}$   &   $3.0\times10^{-4}$     & 9.0        & $3.7\times10^{-3}$   & $2.1\times10^{-4}$  &   0.19\\
3C 317       & 0.0345  & 50   & 8.80   &  $3.4\times10^{-6}$    &  $1.7\times10^{-5}$   &   $1.9\times10^{-4}$     & 8.9        & $4.7\times10^{-3}$   & $4.5\times10^{-3}$  &   0.28\\
B2 0055+30   & 0.0165  & 35   & 9.18   &  $2.4\times10^{-6}$    &  $7.0\times10^{-6}$   &   $2.0\times10^{-4}$     & 3.5        & $2.7\times10^{-3}$   & $1.4\times10^{-2}$  &   0.08\\
3C 66B       & 0.0213  & 45   & 8.84   &  $1.0\times10^{-6}$    &  $1.0\times10^{-5}$   &   $<1.7\times10^{-4}$    & $>5.9$     & $<2.6\times10^{-3}$  & $2.5\times10^{-2}$  &   $>0.18$\\
3C 449       & 0.0171  & 80   & 8.42   &  $8.0\times10^{-7}$    &  $2.0\times10^{-5}$   &   $1.4\times10^{-4}$     & 14.3       & $1.9\times10^{-3}$   & $1.7\times10^{-5}$  &   0.44\\
3C 272.1     & 0.0035  & 63   & 8.35   &  $8.3\times10^{-8}$    &  $4.9\times10^{-6}$   &   $<6.7\times10^{-5}$    & $>7.3$     & $<1.9\times10^{-3}$  & $6.0\times10^{-2}$  &   $>0.22$\\

\enddata
\tablenotetext{a}{inclination angle of the jet with an uncertainty
of several degrees. }
 \tablenotetext{b}{in unit of $\rm M_{\odot}$,
which is derived from the correlation between the stellar velocity
dispersion of the host bulge and its B band magnitude(Marchesini et
al. 2004).} \tablenotetext{c}{$L_{X}$ is the X-ray luminosity in
2-10 keV band.} \tablenotetext{d}{ratio is $\dot{m}_{\rm
jet}/\dot{m}(10\ R_{\rm S})$.} \tablenotetext{e}{$\dot{m}(R_{\rm
B})$ is the dimensionless accretion rate at the Bondi radius through
our spectra fitting.} \tablenotetext{f}{$\dot{m}_{B}$ is the
dimensionless Bondi accretion rate estimated from the X-ray
observation.}
\end{deluxetable}

\clearpage

\begin{figure}
\epsscale{1.0} \plotone{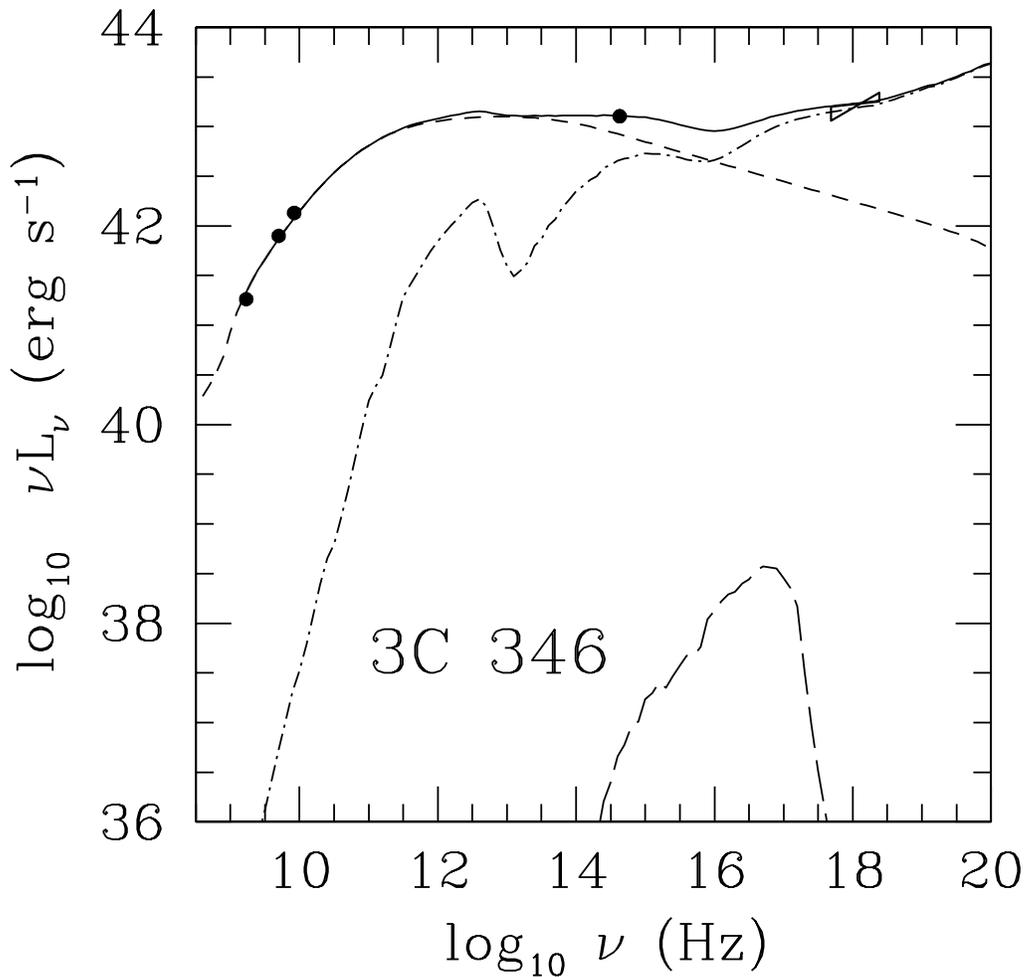} \caption{Spectral modeling results
using the ADAF-jet model for 3C 346. The X-ray luminosity of
this source is $L_{\rm X}= 1.8 \times
10^{-4} L_{\rm Edd}$. The dot-dashed, dashed, and the solid lines
show the emissions from the ADAF, jet, and their sum, respectively.
The thin long-dashed line is synchrotron-self-Compton spectrum of
the jet. For this source, the X-ray emission is dominated by the
ADAF rather than the jet. \label{fig1}}
\end{figure}

\begin{figure}
\epsscale{1.0} \plotone{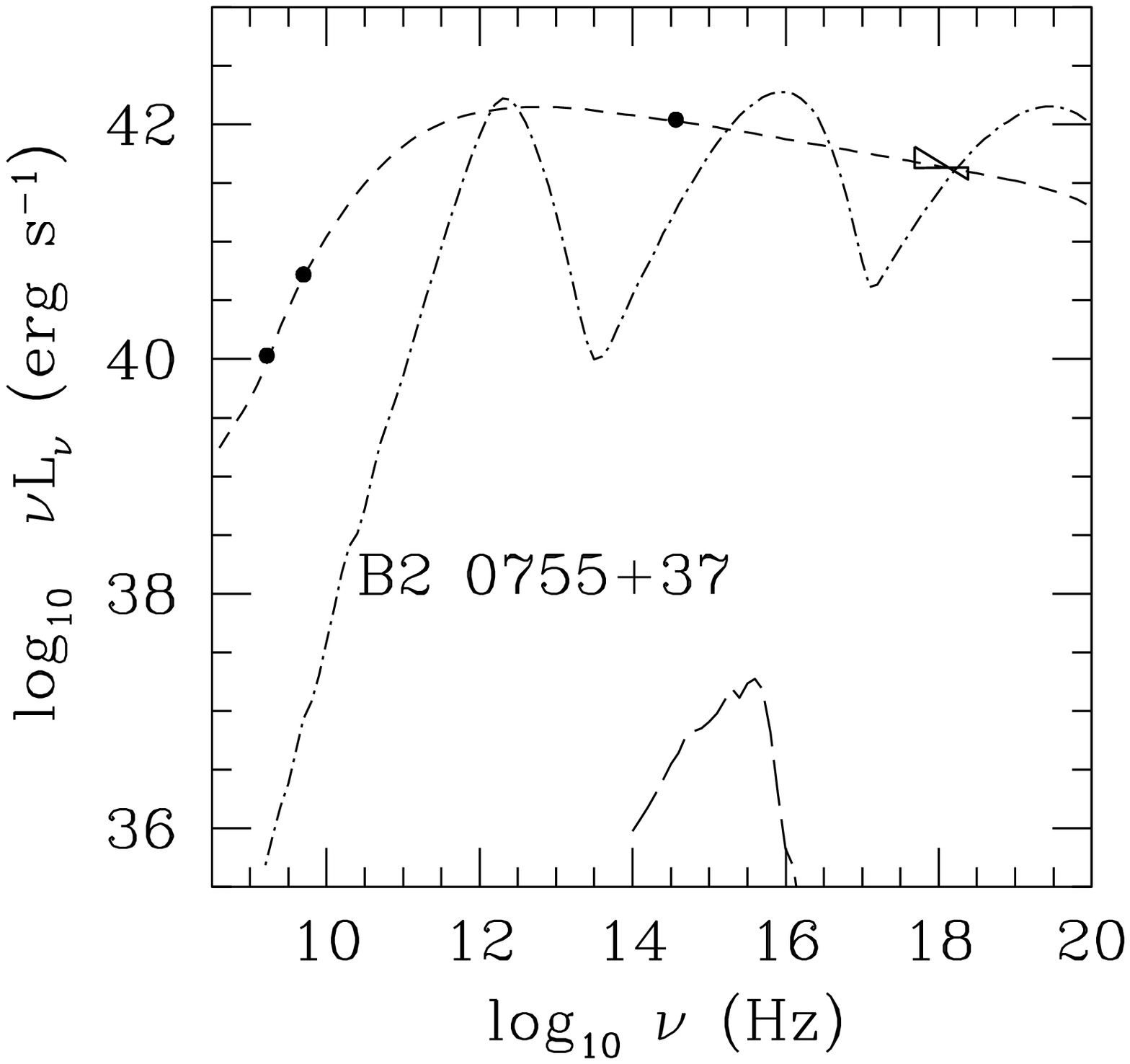} \caption{Spectral modeling results
for B2 0755+37 ($L_{\rm X}= 5.2 \times 10^{-6} L_{\rm Edd}$).
The dashed line shows the emissions from the jet, it explains the
X-ray spectrum very well. Also shown in the figure is the
synchrotron-self-Compton spectrum of the jet (long-dashed line) and
the spectrum produced by an ADAF model (dot-dashed line).
The parameters of the ADAF are chosen so that it can
produce a ''correct'' X-ray flux. Obviously the
ADAF model cannot explain the X-ray spectrum since the spectrum it predicts
is too hard. \label{fig2}}
\end{figure}

\begin{figure} \plotone {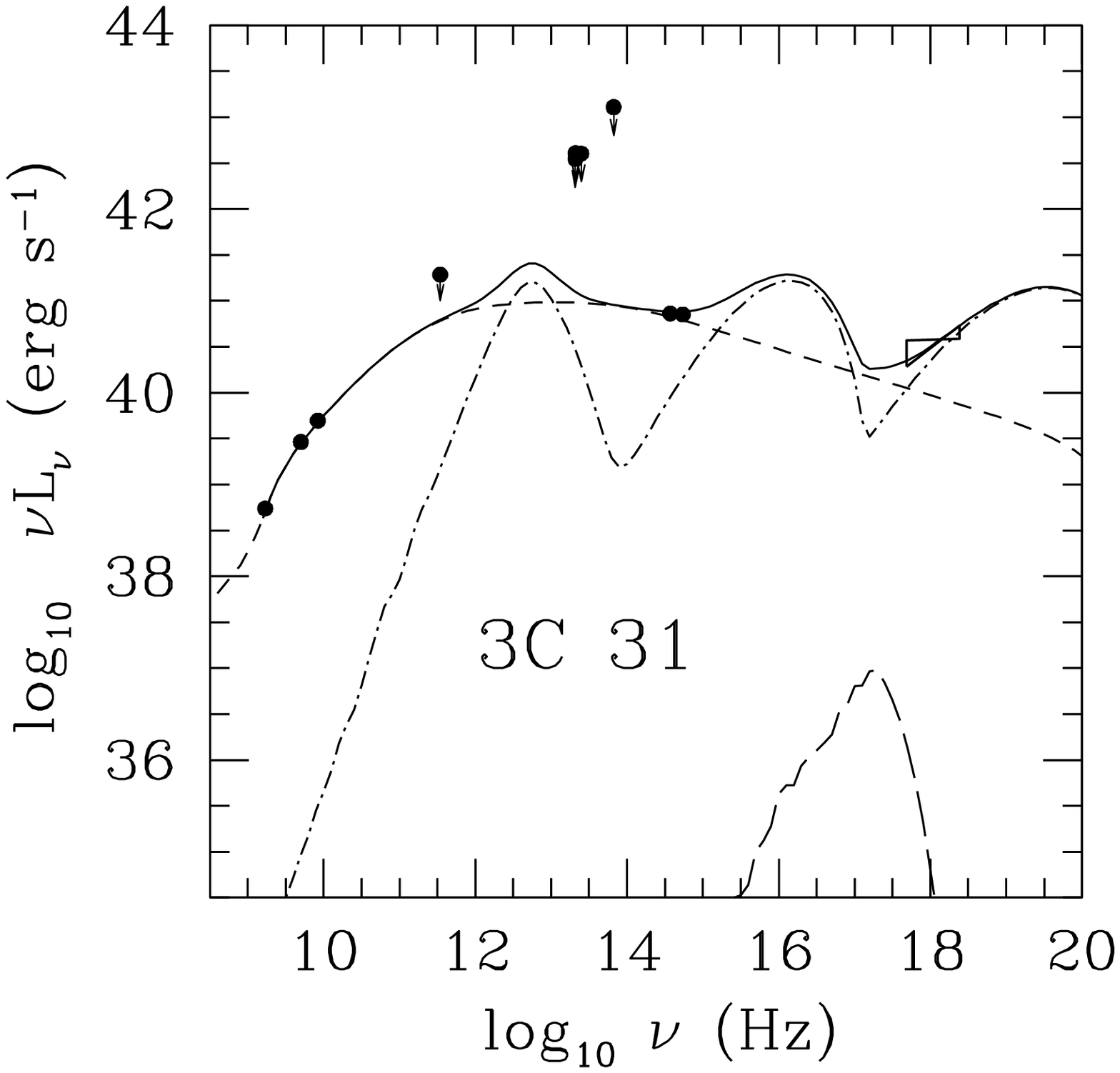}
\caption{Spectral modeling results for 3C 31 ($L_{\rm X}=4.4
\times 10^{-6} L_{\rm Edd}$). The dot-dashed, dashed, and the solid
lines show the emissions from the ADAF, jet, and their sum,
respectively. The long-dashed line is synchrotron-self-Compton
spectrum of the jet. For this source, the contributions from the
ADAF and the jet are comparable.}
\end{figure}

\begin{figure} \plotone {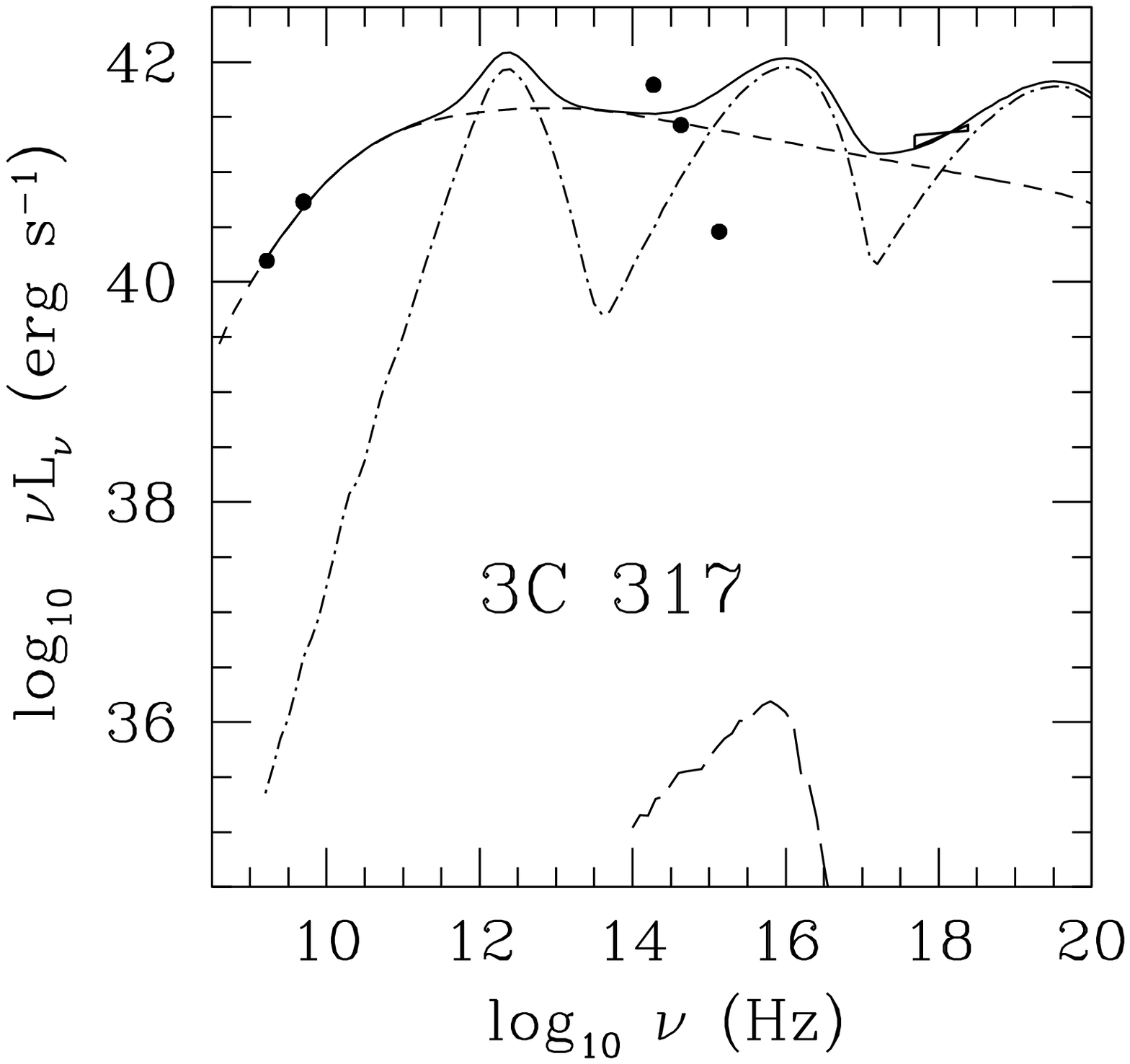}
\caption{Spectral modeling results for 3C 317 ($L_{\rm X}=
3.4 \times 10^{-6} L_{\rm Edd}$). The dot-dashed, dashed, and the
solid lines show the emissions from the ADAF, jet, and their sum,
respectively. The long-dashed line is synchrotron-self-Compton
spectrum of the jet. For this source, the contributions from the
ADAF and the jet are comparable.}
\end{figure}

\begin{figure} \plotone {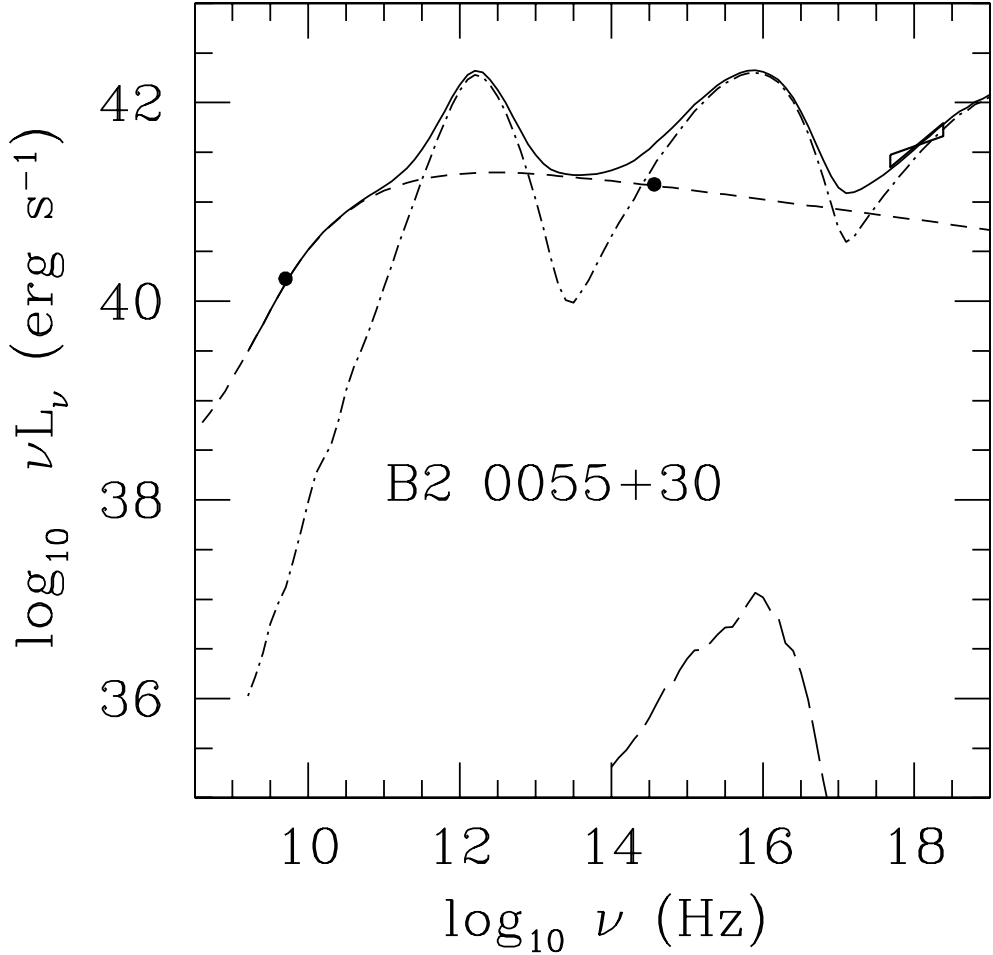}
\caption{Spectral modeling results for B2 0055+30 ($L_{\rm X}=
2.4 \times 10^{-6} L_{\rm Edd}$). The dot-dashed, dashed, and the
solid lines show the emissions from the ADAF, jet, and their sum,
respectively. The long-dashed line is synchrotron-self-Compton
spectrum of the jet. The X-ray emission in this source
is dominated by the ADAF, and the jet contributes a small fraction
in soft X-ray band.}
\end{figure}

\begin{figure} \plotone {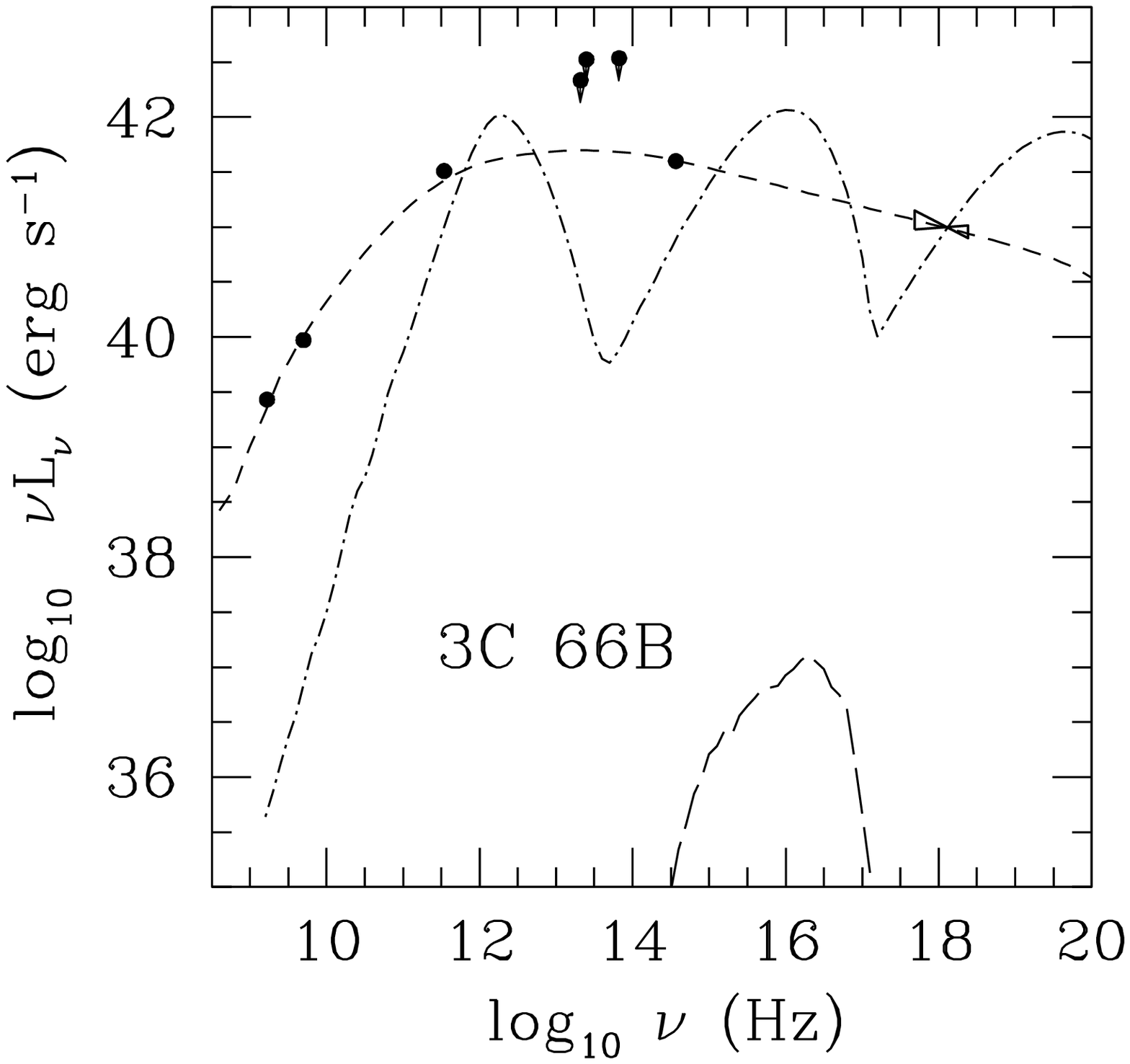}
\caption{Spectral modeling results for 3C 66B ($L_{\rm X}=
1.0 \times 10^{-6} L_{\rm Edd}$).
The dashed line shows the emissions from the jet, it explains the
X-ray spectrum very well. Also shown in the figure is the
synchrotron-self-Compton spectrum of the jet (long-dashed line) and
the spectrum produced by an ADAF model (dot-dashed line).
The parameters of the ADAF are chosen so that it can
produce a ''correct'' X-ray flux. Obviously the
ADAF model cannot explain the X-ray spectrum since the spectrum it predicts
is too hard.}
\end{figure}

\begin{figure} \plotone {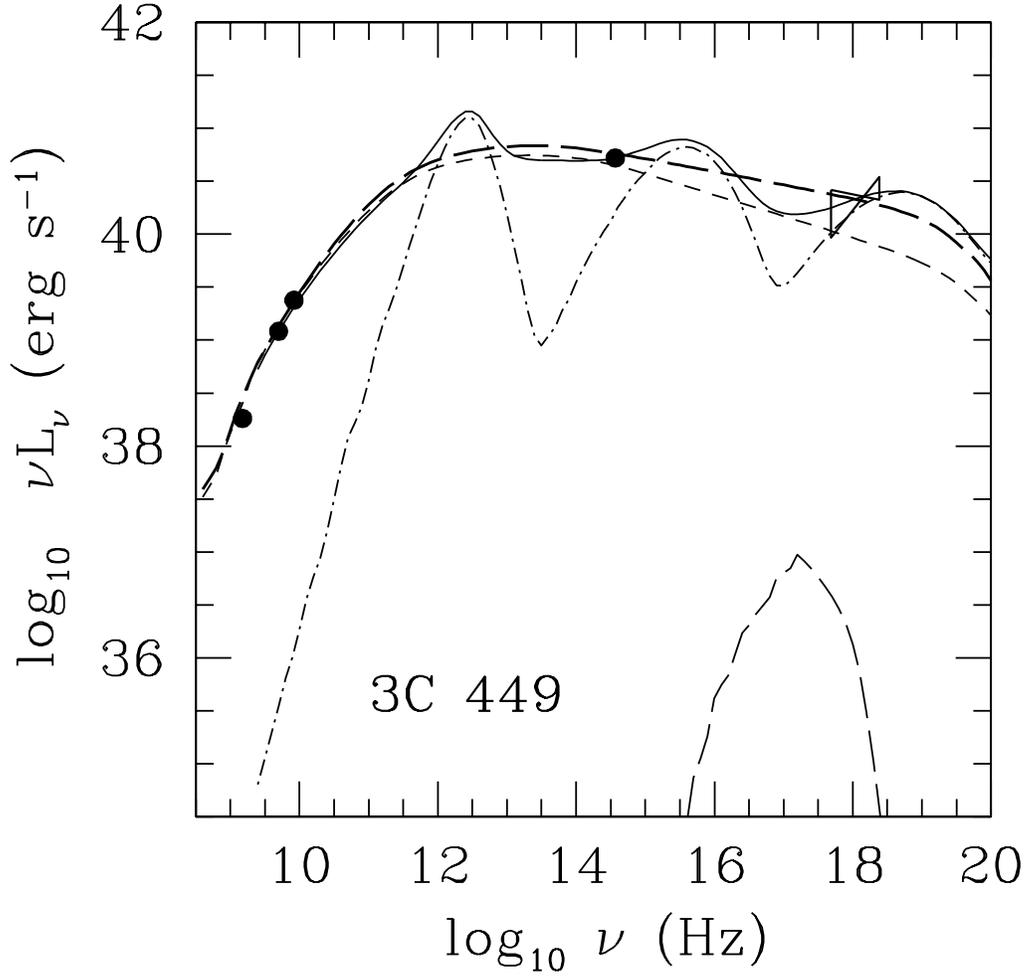}
\caption{Spectral modeling results for 3C 449 ($L_{\rm X}=
8.0 \times 10^{-7} L_{\rm Edd}$). The dot-dashed, thin short-dashed,
and the solid lines show the emissions from the ADAF, jet, and their
sum, respectively. The solid lines can well fit the X-ray spectrum.
For this source, due to the relatively large error bar, the X-ray
spectrum can also be fit by a pure jet model, as shown by the thick
long-dashed line. }
\end{figure}

\begin{figure} \plotone {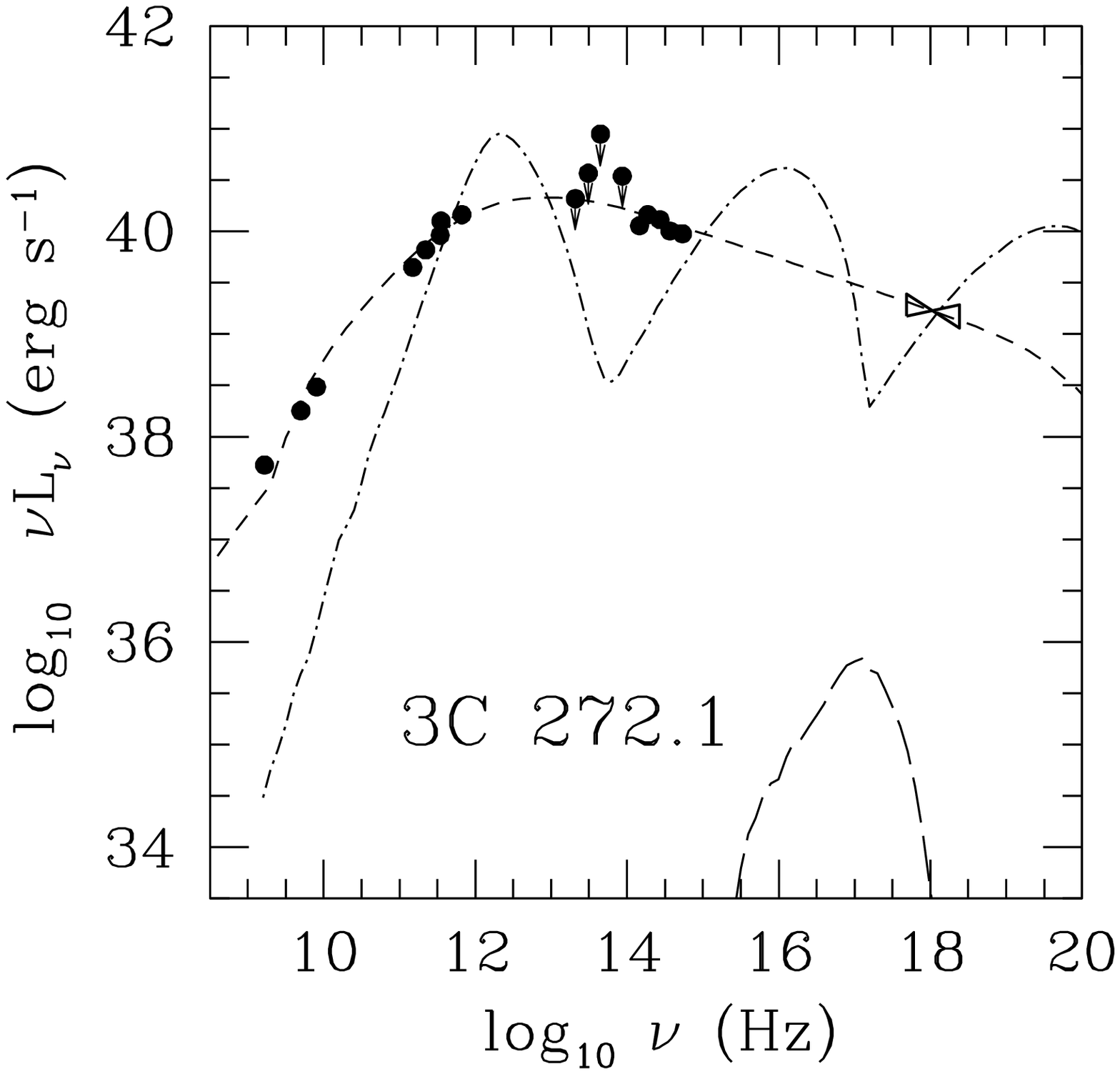}
\caption{Spectral modeling results for 3C 272.1 ($L_{\rm
X}= 6.8 \times 10^{-8} L_{\rm Edd}$).
The dashed line shows the emissions from the jet, it explains the
X-ray spectrum very well. Also shown in the figure is the
synchrotron-self-Compton spectrum of the jet (long-dashed line) and
the spectrum produced by an ADAF model (dot-dashed line).
The parameters of the ADAF
are chosen so that it can produce a ''correct'' X-ray flux. Obviously the
ADAF model cannot explain the X-ray spectrum since the spectrum it predicts
is too hard. }
\end{figure}

\end{document}